\documentclass[runningheads]{llncs}
\usepackage[T1]{fontenc}
\usepackage{graphicx,verbatim}
\usepackage{multirow}
\usepackage{amsmath}
\usepackage{bbding}
\begin{document}
\title{Physics-Informed Implicit Neural Representations for Joint B0 Estimation and Echo Planar Imaging}
\titlerunning{Physics-Informed INRs for Joint B0 Estimation and Echo Planar Imaging}
%
%% Removed for anonymized MICCAI 2025 submission
\author{
Wenqi Huang\inst{1}\Envelope \and
Nan Wang\inst{2} \and
Congyu Liao\inst{2,3} \and
Yimeng Lin\inst{2} \and
Mengze Gao\inst{2} \and \\
Daniel Rueckert\inst{1,4} \and
Kawin Setsompop\inst{2}
}
\authorrunning{W. Huang et al.}
\institute{Technical University of Munich, Germany \and
Stanford University, United States \and
University of California, San Francisco, United States \and
Imperial College London, United Kingdom \\
% \email{wenqi.huang@tum.de}\\
\email{wenqi.huang@tum.de}\\
}

% \author{Anonymized Authors}  %% Added for anonymized MICCAI 2025 submission
% \authorrunning{Anonymized Author et al.}
% \institute{Anonymized Affiliations \\
%     \email{email@anonymized.com}}

\maketitle              % typeset the header of the contribution
\begin{abstract}
% EPI imaging, clinical benifits, B0 inhomogeneity caused distortion, image space correction, INR, physcis informed correction model, joint optimizing of B0 and distortion-free image. results. 
Echo Planar Imaging (EPI) is widely used for its rapid acquisition but suffers from severe geometric distortions due to B0 inhomogeneities, particularly along the phase encoding direction. Existing methods follow a two-step process: reconstructing blip-up/down EPI images, then estimating B0, which can introduce error accumulation and reduce correction accuracy. This is especially problematic in high B0 regions, where distortions align along the same axis, making them harder to disentangle. In this work, we propose a novel approach that integrates Implicit Neural Representations (INRs) with a physics-informed correction model to jointly estimate B0 inhomogeneities and reconstruct distortion-free images from rotated-view EPI acquisitions. INRs offer a flexible, continuous representation that inherently captures complex spatial variations without requiring predefined grid-based field maps. By leveraging this property, our method dynamically adapts to subject-specific B0 variations and improves robustness across different imaging conditions. Experimental results on 180 slices of brain images from three subjects demonstrate that our approach outperforms traditional methods in terms of reconstruction quality and field estimation accuracy. 
\footnote[1]{Code available: 
\url{https://github.com/wenqihuang/PINR}}
\footnote[2]{Accepted at MICCAI 2025.}
% \url{https://github.com/wenqihuang/PINR}}

\keywords{MRI  \and Echo Planar Imaging \and B0 Inhomogeneity \and Implicit Neural Representations \and Distortion Correction.}
% Authors must provide keywords and are not allowed to remove this Keyword section.

\end{abstract}
\section{Introduction}
Magnetic Resonance Imaging (MRI) is widely used in clinical and research settings for non-invasive imaging with excellent soft tissue contrast. Echo Planar Imaging (EPI), a commonly used fast MRI technique, enables rapid acquisition of images, making it particularly advantageous for functional and diffusion imaging~\cite{stehling1991echo,setsompop2012improving}. However, its high-speed acquisition causes geometric distortions along the phase encoding direction due to B0 field inhomogeneities, leading to stretching, compression, intensity modulation, and misalignment, especially near air-tissue interfaces~\cite{andersson2003correct,smith2004advances}. These artifacts compromise spatial accuracy, affecting quantitative analysis and clinical diagnosis.

Several techniques have been developed to correct B0-induced distortions in EPI. Some use pre-calculated field maps acquired before the scan, but these require extra scan time and may suffer from physiological B0 fluctuations (e.g., breathing, cardiac pulsation, motion) and error accumulation~\cite{zeng2002image}. 
Others, such as blip-up/down acquisitions, estimate B0 by capturing images with opposite phase encoding gradients~\cite{andersson2003correct,smith2004advances,andersson2016integrated,liao2019highly,liao2021distortion}. The distortions in the images with opposite phase encoding appear also in opposite directions, allowing B0 estimation through displacement analysis. While avoiding pre-scanning, this approach captures distortions along a single axis, making it difficult to resolve complex B0 variations, especially in high-susceptibility regions. Some studies have explored multi-shot EPI~\cite{bhushan2014improved,mani2017multi}, but all of the existing methods are restricted to orthogonal directions. Acquiring phase encoding in rotated directions can capture more information of image distortion in multiple orientations, but the reconstruction and post-processing of off-grid data is challenging with conventional discrete representations. They could also encounter accumulated errors, as any inaccuracies in distorted image reconstruction, B0 estimation, propagate into the reconstructed image.

% Implicit neural representations (INRs) have recently gained significant attention in computer vision and medical imaging, demonstrating their capability to model complex spatial variations in a continuous and high-fidelity manner. Unlike conventional grid-based representations, INRs encode signals as continuous functions, allowing them to effectively capture fine details without being constrained by a fixed resolution. These properties have led to their successful application in CT reconstruction\cite{reed2021dynamic}, MRI reconstruction\cite{shen2022nerp,huang2023neural,spieker2023iconik,kunz2024implicit,feng2025spatiotemporal}, and other medical imaging tasks. Given their powerful representational capabilities, INRs provide a promising approach for B0 field estimation and image reconstruction in EPI. 
Implicit neural representations (INRs) have gained significant attention in computer vision and medical imaging for modeling complex signals. Unlike grid-based methods, INRs encode signals as continuous functions, capturing fine details without fixed resolution constraints. They have been successfully applied to CT and MRI reconstruction~\cite{reed2021dynamic,shen2022nerp,huang2023neural,spieker2023iconik,kunz2024implicit,feng2025spatiotemporal}, making them a promising approach for B0 estimation and image reconstruction in EPI.

In this work, we propose a novel joint B0 estimation and image reconstruction framework that utilizes two INR networks, one for modeling the B0 field and another for the distortion-free image. Our method integrates an MR physics-informed forward model that warps the distortion-free image into distorted EPI images, which are then compared with multi-coil rotated-view EPI acquisitions to supervise the training of both networks. The rotated EPIs distribute distortions across multiple orientations, facilitating the disentanglement and correction of distortions. Due to the continuous representation capability of INRs, rotation is handled directly in the coordinate space rather than requiring interpolation in the image space. By jointly optimizing B0 estimation and image reconstruction, our approach mitigates error accumulation and enhances robustness across diverse imaging conditions. Our key contributions are as follows:
\begin{enumerate}
\item We propose a novel \textbf{P}hysics-informed \textbf{I}mplicit \textbf{N}eural \textbf{R}epresentation B0 estimation and image reconstruction framework, termed as PINR, for rotated multi-view EPI. 
%As far as we know, this is the first work using INRs to help solving the field imperfection problem in MRI.
We realized joint parallel imaging reconstruction across multi-views at the same time as estimating B0 field inhomogeneity all in one go rather than separately, and no need for interpolation thanks to the continuous representation capability of INRs.
\item We leverage rotated-view EPI acquisitions to improve B0 field estimation by distributing distortions across multiple orientations, facilitating better disentanglement and correction. 
\item Our approach eliminates the need for additional calibration scans, making the method more time-efficient while maintaining high reconstruction accuracy. By integrating INR-based B0 estimation directly into the reconstruction process through a physics-informed forward model, we reduce error accumulation and enhance robustness across diverse imaging conditions.
\item We evaluated our method on high-resolution MRI simulated rotated-view EPI data and prospectively sampled blip-up/down acquisitions, both shows superier image reconstruction and B0 estimation comparing to state-of-the-art methods.

\end{enumerate}

\section{Methods}
\subsection{EPI and B0 Inhomogeneity}
% MRI reconstruction, with b0 imperfection, with multi-view EPIs.
EPI achieves fast acquisition through its unique single-shot or multi-shot echo train sampling trajectory. Unlike Cartesian sequences that segment \textit{k}-space acquisition, EPI samples \textit{k}-space in a zig-zag pattern using alternating readout gradients. With a homogeneous B0 field, the acquired \textit{k}-space directly corresponds to a distortion-free image after inverse Fourier transformation. However, B0 inhomogeneity introduces an additional phase term at each \textit{k}-space point, dependent on local field variations. Mathematically, the acquired EPI \textit{k}-space data. Mathematically, the acquired EPI \textit{k}-space data \( d(k_x, k_y, t) \) is expressed as: 
% \begin{equation}
% S(k_x, k_y, t) = \int_{\Omega} m(x, y) e^{-i (k_x x + k_y y)} e^{-i \gamma B_0(x,y) t} dx dy,
% \end{equation}
\begin{equation}
d(k_x, k_y, t) = \int_{\Omega} \mathcal{S}(x,y)m(x, y) e^{-i (k_x x + k_y y)} e^{-i \gamma B_0(x,y) t} dx dy,
\end{equation}
here \( m(x, y) \) is the true image intensity, \(\mathcal{S}(x,y)\) represents coil sensitivity maps, \( \gamma \) is the gyromagnetic ratio, and \( B_0(x,y) \) represents the B0 field inhomogeneity at location \( (x, y) \). The integration domain \( \Omega \) represents the spatial region over which the MRI signal is acquired, encompassing all imaged tissue areas. The undesired phase term \( e^{-i \gamma B_0(x,y) t} \) accrual alters the spatial encoding, leading to geometric distortions and signal misregistration in the reconstructed image, particularly along the phase encoding direction. 
\begin{figure}[!t]
    \centering
    \includegraphics[width=\linewidth]{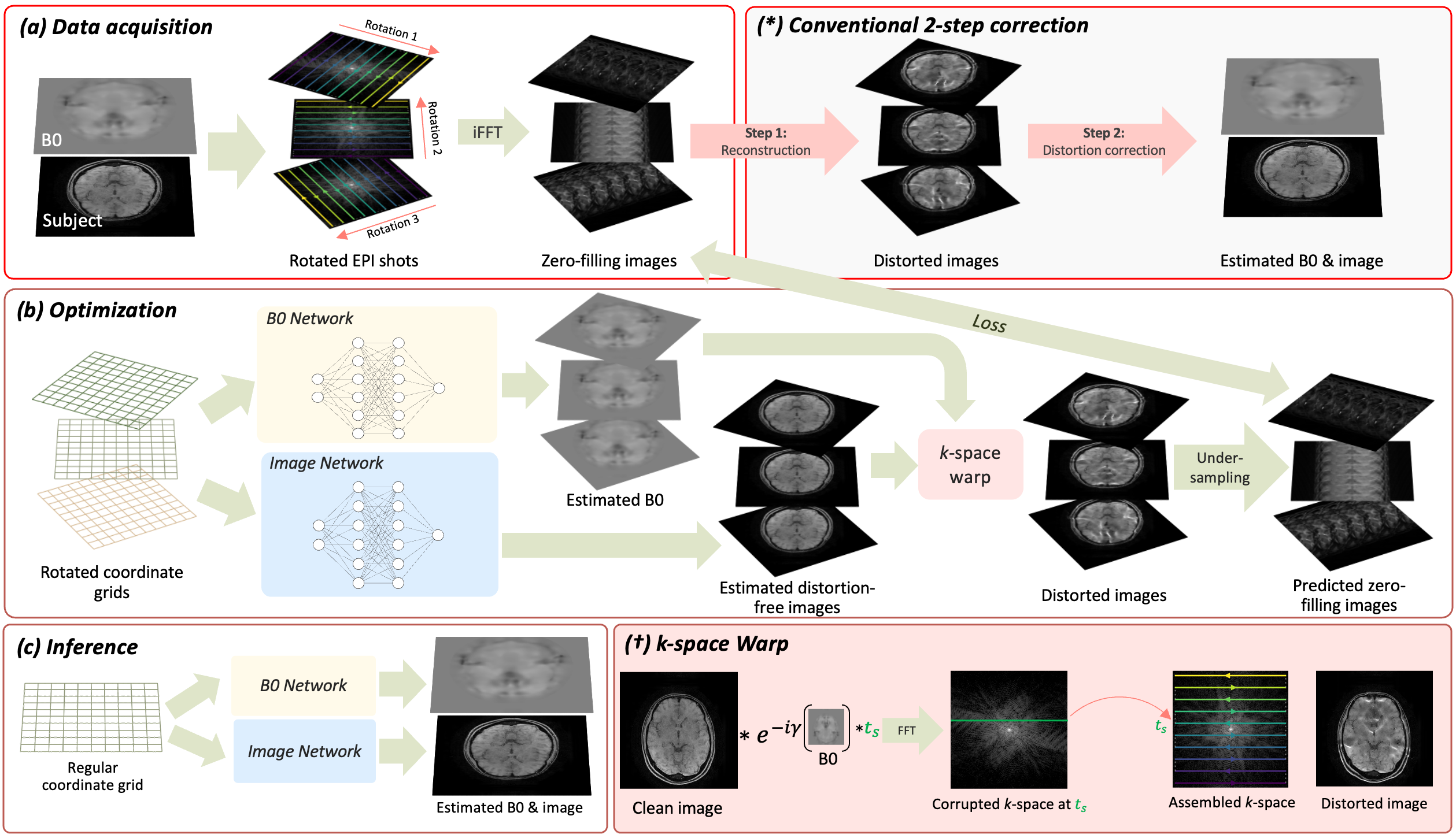}
    \caption{Overview of the proposed PINR. \textbf{(a) Rotated-view EPI acquisition:} Acquiring EPI images at different phase encoding directions distributes distortions and aliasing due to B0 inhomogeneity and \textit{k}-space undersampling. \textbf{(*) Existing methods} use two views (blip-up/down) to reconstruct a distorted image, then estimate B0 for correction. \textbf{(b) Optimization:} Our method inputs rotated coordinate grids into INR networks for B0 and distortion-free image estimation, enforcing MRI physics for training. \textit{k}-space warping is indicated by \textbf{(†)}. \textbf{(c) Inference:} A regular spatial grid is fed into trained INR networks to obtain the B0 field and distortion-free image.}
    \label{fig:main}
\end{figure}

Correcting distortions requires estimating the underlying B0 field \( B_0(x,y) \). Traditional approaches involve acquiring field maps or using blip-up/blip-down acquisitions, but face limitations in scan time and resolving distortions in highly inhomogeneous regions. We propose to use rotated-view EPI acquisitions, which distribute distortions across multiple orientations, allowing for improved B0 estimation and correction. Let \( \theta_n \) represent the phase encoding direction for the \( n \)-th rotated EPI acquisition, with \(N\) total rotations. 
%The observed \textit{k}-space signal in the presence of B0 inhomogeneity can be expressed as:
% \begin{equation}
% S_n(k_x, k_y, t) = \int_{\Omega} m(x, y) e^{-i (k_x x + k_y y)} e^{-i \gamma B_0(x,y) t} dx dy,
% \end{equation}
The B0 field estimation and image reconstruction problem can then be formulated as a joint optimization across multiple rotated views:
% \begin{equation}
% \widehat{B_0}, \hat{m} = \arg\min_{B_0, m} \sum_{n=1}^N \| d_n(k_x, k_y) - \mathcal{A}_{n}(m(x, y) e^{-i \gamma B_0(x,y) t_s}) \|^2 + \mathcal{R}(B_0, m),
% \label{eq:jointopt}
% \end{equation}

\begin{equation}
\widehat{B_0}, \hat{m} = \arg\min_{B_0, m} \sum_{n=1}^N \| d_n(k_x, k_y) - \sum_s\Lambda_{n,s}\mathcal{A}_{n}(m(x, y) e^{-i \gamma B_0(x,y) t_s}) \|^2 + \mathcal{R}(B_0, m),
\label{eq:jointopt}
\end{equation}
where \(\mathcal{A}_n=\mathcal{M}_n\mathcal{FS}\) represents the forward operator of MRI incorporating the coil sensitivity maps \(\mathcal{S}\), the Fourier transform \(\mathcal{F}\), and the undersampling mask \(\mathcal{M}_n\) corresponding to the \(n\)-th rotated acquisition. 
% \(t_s\) is the echo time of segment \(s\), segment selection mask \( \Lambda_{m,s}\) picks the corresponding \textit{k}-space segments and summing across \(s\) to get a full \textit{k}-space.
\( t_s \) denotes the echo time of segment \( s \), while the selection mask \( \Lambda_{n,s} \) extracts \textit{k}-space segments, summing over \( s \) to reconstruct the full \textit{k}-space.
The term \(\mathcal{R}(B_0, m)\) acts as a regularization to enforce prior constraints on \(B_0\) and \(m\). When \(N=2\) and \(|\theta_1-\theta_2|=180\)°, this formulation reduces to the standard joint B0 estimation and image reconstruction problem for blip-up/down EPI acquisitions. 

\subsection{Implicit Neural Representations for Rotated-view EPI} 
% Given these challenges, traditional methods struggle to jointly estimate \( B_0(x,y) \) and reconstruct \( m(x,y) \) due to the problem’s highly non-linear nature. Additionally, the reliance on grid-based processing has restricted exploration to four primary directions (up, down, left, right), limiting the flexibility of EPI acquisitions. When only two opposing phase encoding directions are used—a common practice—the ability to fully resolve distortions is further constrained by the limited directional information available. To address these limitations, we propose leveraging Implicit Neural Representations (INRs) to model both the image and the B0 field as continuous functions, enabling a more flexible and effective reconstruction framework. INRs are a group of coordinate-based deep learning models that map spatial coordinates (e.g. \((x,y)\)) to signal values (e.g. image intensity) through a neural network. Unlike traditional grid-based representations, INRs model signals as continuous functions, allowing them to naturally capture fine details without being constrained by discrete pixel grids. This property makes INRs particularly well-suited for rotated view EPI. We propose using an INR to model the image \( m(x,y) \) and the B0 field \( B_0(x,y) \) as:
% Given these challenges, traditional methods struggle to simultaneously estimate \( B_0(x,y) \) and reconstruct \( m(x,y) \) due to the problem’s highly non-linear nature. 
Typical approaches involve sequentially estimating B0 before jointly reconstructing the \textit{k}-space data. This limits the acceleration of each EPI shot, as highly undersampled acquisitions make it difficult to estimate a reliable B0 field upfront. Additionally, the reliance on grid-based processing has restricted exploration to four primary directions (up, down, left, right), limiting the flexibility of EPI acquisitions. When only two opposing phase encoding directions are used, resolving distortions is further constrained by limited directional information. To overcome these limitations, we leverage INRs to model both the image and B0 field as continuous functions, enabling a more flexible and accurate reconstruction framework. INRs are coordinate-based deep learning models that map spatial coordinates (e.g., \((x,y)\)) to signal values (e.g., image intensity) through a neural network. Unlike traditional grid-based representations, INRs provide continuous modeling, naturally capturing fine details without pixel grid constraints. This property makes them particularly well-suited for rotated-view EPI. We propose using an INR to represent both the image \( m(x,y) \) and the B0 field \( B_0(x,y) \) as:

\begin{equation}
    m(x,y) = f_\psi(x,y), \quad B_0(x,y) = g_\phi(x,y),
    \label{eq:inrmodels}
\end{equation}
where \( f_\psi \) and \( g_\phi \) are neural networks parameterized by \( \psi \) and \( \phi \), respectively. These networks take spatial coordinates as input and output the corresponding intensity and field values. A key advantage of INRs is their ability to handle rotated-view acquisitions without interpolation. Instead of rotating the image or B0 field directly, INRs allow us to simply rotate the coordinate grid and query the function at the new locations:
\begin{equation}
    \begin{bmatrix} x' \\ y' \end{bmatrix} = \begin{bmatrix} \cos\theta_n & -\sin\theta_n \\ \sin\theta_n & \cos\theta_n \end{bmatrix} \begin{bmatrix} x \\ y \end{bmatrix}.
\end{equation}
This property ensures that rotated views can be handled seamlessly without introducing interpolation artifacts, thereby preserving spatial fidelity.

By incorporating INRs into our joint B0 estimation and image reconstruction framework, we gain a flexible and high-fidelity approach to modeling spatial variations. The continuous representation allows our method to generalize across different imaging conditions while maintaining robustness to distortions.

\subsection{Model Optimization and Inference}
To jointly estimate the B0 field and reconstruct the distortion-free image, we optimize the parameters \( \psi \) and \( \phi \) of the INR networks \( f_\psi(x,y) \) and \( g_\phi(x,y) \). By incorporating these networks into the joint optimization framework in Eq.~\eqref{eq:jointopt}, the learning objective becomes: 
% \begin{equation}
% \widehat{\psi}, \widehat{\phi} = \arg\min_{\psi, \phi} \sum_{n=1}^{N} \| S_n(k_x, k_y) - \mathcal{A}_n(f_\psi(x, y) e^{-i \gamma g_\phi(x,y) t_s}) \|^2 + \mathcal{R}(f_\psi, g_\phi),
% \end{equation}
\begin{equation}
\widehat{\psi}, \widehat{\phi} = \arg\min_{\psi, \phi} \sum_{n=1}^{N} \| \mathcal{A}_n^Hd_n(k_x, k_y) - \mathcal{A}_n^H\sum_s\Lambda_{n,s}\mathcal{A}_n(f_\psi(x, y) e^{-i \gamma g_\phi(x,y) t_s}) \|_{sl1} + \mathcal{R}(f_\psi, g_\phi),
\end{equation}
where \( \mathcal{A}_n \) is the MRI forward operator incorporating sampling, Fourier transform, and coil sensitivity maps for the \(n\)-th view, \( \mathcal{A}_n^H \) is its adjoint operator that convert \textit{k}-space to zero-filled images. The inclusion of \( \mathcal{A}_n^H \) allows loss calculation in image space, mitigating issues caused by the large intensity range of \textit{k}-space values. We choose smoothed L1 loss \(\|\cdot\|_{sl1}\) for the data consistency to prevent exploding gradients like \cite{girshick2015fast}. \( \mathcal{R}(f_\psi, g_\phi) \) serves as a regularization term to enforce smoothness and prior constraints. The process is illustrated in Fig.~\ref{fig:main}b.

Once the models are well-trained, B0 and the distortion-free image can be obtained by querying the INR networks over a regular grid. Specifically, by passing a uniform spatial coordinate grid \( (x,y) \) into \( g_\phi(x,y) \), we obtain the estimated B0 field \( \widehat{B_0} \). Similarly, by querying \( f_\psi(x,y) \), we obtain the final high-fidelity distortion-free image \( \hat{m} \) (see Fig.~\ref{fig:main}c):

\begin{equation}
\widehat{B_0}(x,y) = g_{\widehat{\phi}}(x,y), \quad \hat{m}(x,y) = f_{\widehat{\psi}}(x,y).
\end{equation}

% This inference process is illustrated in Fig.~\ref{fig:main} (b) and (c), where the trained INR networks produce continuous representations of the B0 field and the final corrected image. The continuous nature of INRs enables high-resolution reconstruction without relying on discrete grid-based interpolation, improving both accuracy and robustness.
\section{Experimental Setup}
\subsubsection{Dataset} 
We simulated distorted EPI data at different rotated angles using undistorted images from an accelerated multi-echo GRE dataset acquired with EPTI~\cite{wang2019echo}. The B0 map was estimated from multi-echo images~\cite{wang2025spherical}, and coil sensitivity was obtained using ESPIRiT~\cite{uecker2014espirit}. Data were collected from three healthy volunteers on a 3T scanner (UHP, GE Healthcare) with a \(240 \times 216 \times 240\) mm\(^3\) FOV, 1-mm isotropic resolution, 48 echoes (1 ms echo spacing, TE: 5–53 ms, TR: 60 ms), and a 45-second acquisition time (\( R = 48 \)), with ethical consent. Each scan included 90 brain slices, totaling 270 slices. Distorted images were simulated with a 0.25 ms echo spacing for fully sampled data and \( R \times 0.25 \) ms for undersampling. Two datasets were generated: (1) A two-view set (\( R=4 \), 0° and 180°) simulating blip-up/down, and (2) A three-view set (\( R=6 \), 0°, 120°, 240°) for rotated-view EPI. Ground truth images in the three-view set were interpolated onto a rotated grid, introducing potential systematic errors.

\subsubsection{Model Details}
% The proposed INR framework consists of a B0 network and a image network. Both networks are based on multi-layer perceptrons (MLPs) with hash-grid encoding~\cite{muller2022instant}. The hash-grid encoding uses a hashmap of size 20 with 16 levels, 2 features per level, a base resolution of 16, and a per-level scale of 1.19. The MLPs have 2 hidden layers, each containing 256 nodes. The B0 network's output dimension is 1, and the image network generate an output dimension of 2, which represents the concatenation of the real and imaginary parts of the distortion-free image.
% The model was trained using a smoothed L1 loss to enforce data consistency between the predicted and acquired zero-filling images. To encourage smoothness in the B0 field estimation, we applied total variation (TV) regularization with an initial weighting factor of \( 1 \times 10^{-5} \), which decayed by a factor of 0.1 every 1000 iterations. Regularization was removed in the final 1000 iterations to fully leverage the flexibility of implicit neural representations. The model was optimized using the AdamW optimizer with a learning rate of \( 3 \times 10^{-3} \) for 6000 iterations. Implementation was done in PyTorch and training was performed on an Nvidia A6000 GPU, taking approximately 4 minutes per slice.
The B0 network and image network are both multi-layer perceptrons (MLPs) with hash-grid encoding~\cite{muller2022instant}. The encoding uses a hashmap of size 20 with 16 levels, 2 features per level, a base resolution of 16, and a per-level scale of 1.19. Each MLP has 2 hidden layers with 256 nodes. The B0 network outputs a scalar, while the image network outputs two channels representing the real and imaginary parts of the distortion-free image. The model was trained with a smoothed L1 loss for data consistency and total variation (TV) regularization to enforce smooth B0 estimation. The TV weight started at \( 1 \times 10^{-5} \), decaying by 0.1 every 1000 iterations, and was removed in the last 1000 iterations to fully utilize the INR flexibility. Training used AdamW with a \( 3 \times 10^{-3} \) learning rate for 6000 iterations. 
Our proposed method implemented in PyTorch on an NVIDIA A6000 GPU took around 4 minutes per slice. 

\subsubsection{Evaluation Settings}
% We evaluated our method against the state-of-the-art TOPUP method from FSL \cite{andersson2003correct,smith2004advances}. GRAPPA was used to reconstruct blip-up/down images before applying TOPUP. As TOPUP does not supports rotated acquisitions, we designed two experiments: (1) Comparing B0 estimation and image correction using two EPI views (blip-up/down) to assess the effectiveness of our method. (2) Evaluating our method with three-view EPI acquisitions and comparing the results with both blip-up/down TOPUP and our two-view method. To ensure fair \textit{k}-space coverage, we set \( R=4 \) for two-view experiments and \( R=6 \) for three-view experiments, keeping the total number of acquired \textit{k}-space echo lines equal. For quantitative evaluation, we computed NRMSE, PSNR, and SSIM for images, and absolute difference for B0 estimation.
We evaluated our method against the state-of-the-art TOPUP method from FSL~\cite{andersson2003correct,smith2004advances}, using GRAPPA~\cite{griswold2002grappa} to reconstruct blip-up/down images before correction. As TOPUP does not support rotated acquisitions, we conducted two experiments: (1) Comparing B0 estimation and image correction using two EPI views (blip-up/down) to assess our method’s effectiveness, and (2) Evaluating our method with three-view EPI and comparing it to both TOPUP and our two-view method. To ensure fair \textit{k}-space coverage, we set \( R=4 \) for two-view and \( R=6 \) for three-view experiments, maintaining the same total acquired \textit{k}-space lines. Quantitative evaluation included NRMSE, PSNR, and SSIM for images, and absolute difference for B0 estimation. The TOPUP method processed a full volume of 90 slices in about 10 hours for this high-resolution data.

\begin{table}[h]
    \centering
    \caption{Comparison of Image and B0 Metrics (Mean ± Std)}
    \begin{tabular}{c|ccc|c}
        \hline
        \multirow{2}{*}{\textbf{Method}} & \multicolumn{3}{c|}{\textbf{Image Metrics}} & \textbf{B0 Metrics} \\
        \cline{2-5}
         & NRMSE $\downarrow$ & PSNR (dB) $\uparrow$ & SSIM $\uparrow$ & Diff. (Hz) $\downarrow$ \\
        \hline
        TOPUP & 0.0534 ± 0.0153 & 37.1972 ± 2.0703 & 0.9666 ± 0.0171 & 3.0505 ± 1.3535 \\
        PINR-2v & \textbf{0.0143} ± 0.0169 & \textbf{51.9853} ± 7.1921 & \textbf{0.9955} ± 0.0054 & \textbf{0.6017} ± 0.3742 \\
        PINR-3v & 0.0441 ± 0.0104 & 38.7221 ± 1.8140 & 0.9805 ± 0.0097 & 1.0657 ± 0.4628 \\
        \hline
    \end{tabular}
    \label{tab:metrics_comparison}
\end{table}

\section{Results}
% Table~\ref{tab:metrics_comparison} summarizes the mean and standard deviation of the quantitative metrics for TOPUP, our proposed PINR method with two views (PINR-2v) and three views (PINR-3v). Our method outperforms TOPUP across all metrics in both the two-view and three-view cases. Notably, for three-view PINR, the image quality metrics and B0 error appear slightly worse than the two-view case. We attribute this to interpolation artifacts introduced during multi-view image generation in data preparation. Despite this interpolation error in the training data, PINR still achieves better results than the two-step, blip-up/down and image-domain-based TOPUP method.
Table~\ref{tab:metrics_comparison} summarizes the mean and standard deviation of quantitative metrics for TOPUP, PINR with two views (PINR-2v), and three views (PINR-3v). Our method outperforms TOPUP across all metrics. While PINR-3v shows slightly higher B0 error and lower image quality metrics than PINR-2v, likely due to interpolation artifacts in multi-view image generation, it still surpasses the two-step, image-domain-based TOPUP method.

\begin{figure}[h]
    \centering
    \includegraphics[width=0.95\linewidth]{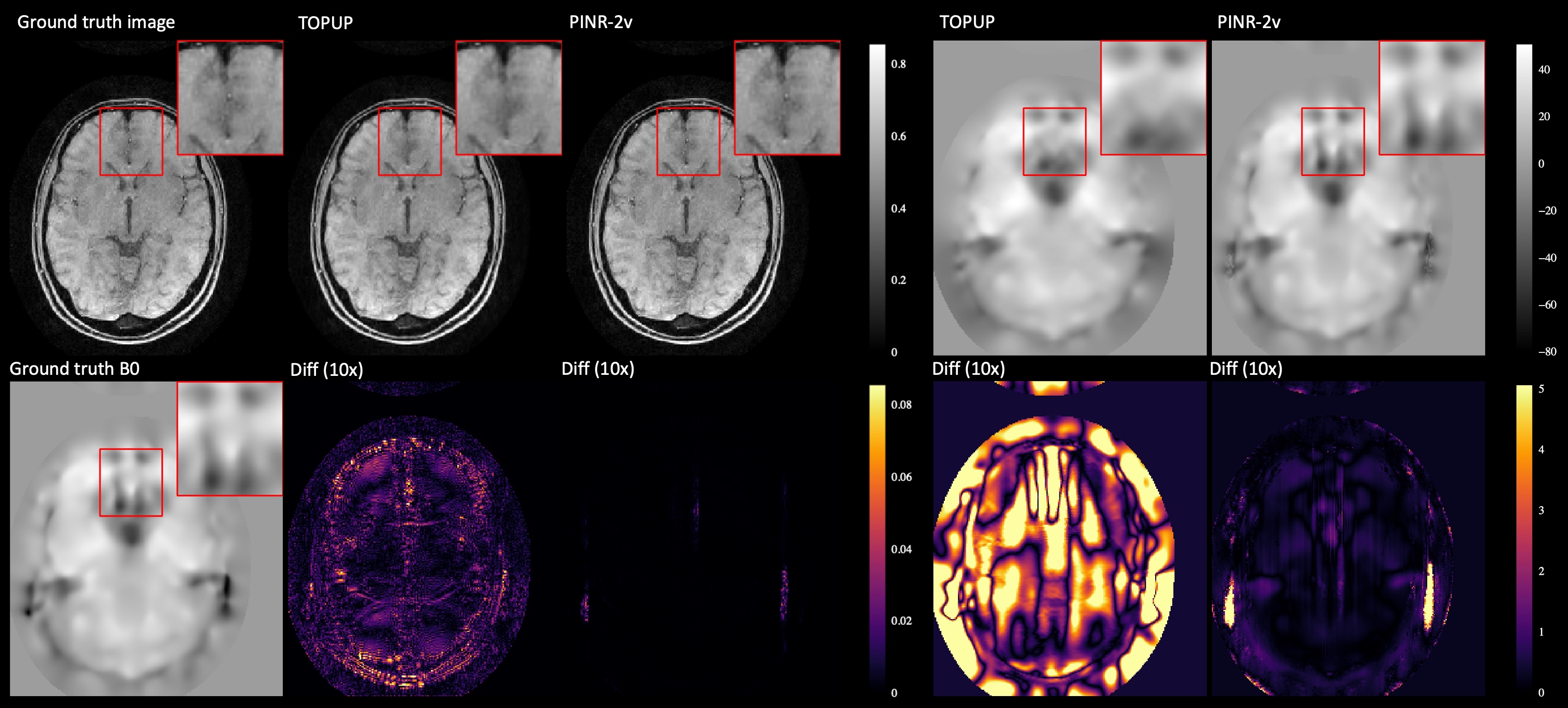}
    \caption{Comparison with two-view (blip-up/down) acquisition. The first column shows the ground truth image and B0 field. The second and third columns present TOPUP-corrected results and our method’s results, along with their error maps. The right two columns display B0 estimates from TOPUP and our method, with corresponding error maps below. The red box highlights the zoomed-in region of interest (ROI).}
    \label{fig:comparison_2view}
\end{figure}
% Figure~\ref{fig:comparison_2view} compares the results of TOPUP and our two-view PI2NR method. By examining the ROI in the image columns, we observe that TOPUP produces a blurrier output, while our method preserves finer structural details. In the error maps, TOPUP results contain residual aliasing artifacts, likely caused by errors in the GRAPPA reconstruction step. Since TOPUP operates solely in the image domain, and its image reconstruction and b0 correction steps are seperated, any imperfections in the distorted EPI reconstruction process propagate into B0 estimation, ultimately affecting the corrected image. Looking at the B0 estimation results, our method consistently exhibits lower error across the entire field. In the ROI, PI2NR captures finer variations in B0, benefiting from the superior representational flexibility of INRs and the joint optimization scheme. In contrast, TOPUP's B0 estimation is constrained by the assumptions of a Gaussian model, leading to oversmoothing and reduced accuracy in regions with strong susceptibility variations. 
% These findings demonstrate the advantage of our joint optimization framework in achieving more precise B0 estimation and distortion correction.

Figure~\ref{fig:comparison_2view} compares TOPUP and our two-view PINR. The ROI shows that TOPUP produces blurrier images, while PINR preserves finer structures. Error maps reveal residual aliasing in TOPUP, likely from GRAPPA reconstruction errors. Since TOPUP operates in the image domain with separate reconstruction and B0 correction steps, errors in distorted EPI propagate into B0 estimation, degrading the final image. PINR consistently achieves lower B0 error, particularly in the ROI, benefiting from INRs' flexibility and joint optimization. In contrast, TOPUP's Gaussian-based B0 estimation oversmooths high-susceptibility regions. These results highlight the advantage of our joint framework for precise B0 estimation and distortion correction.

\begin{figure}[h]
    \centering
    \includegraphics[width=0.95\linewidth]{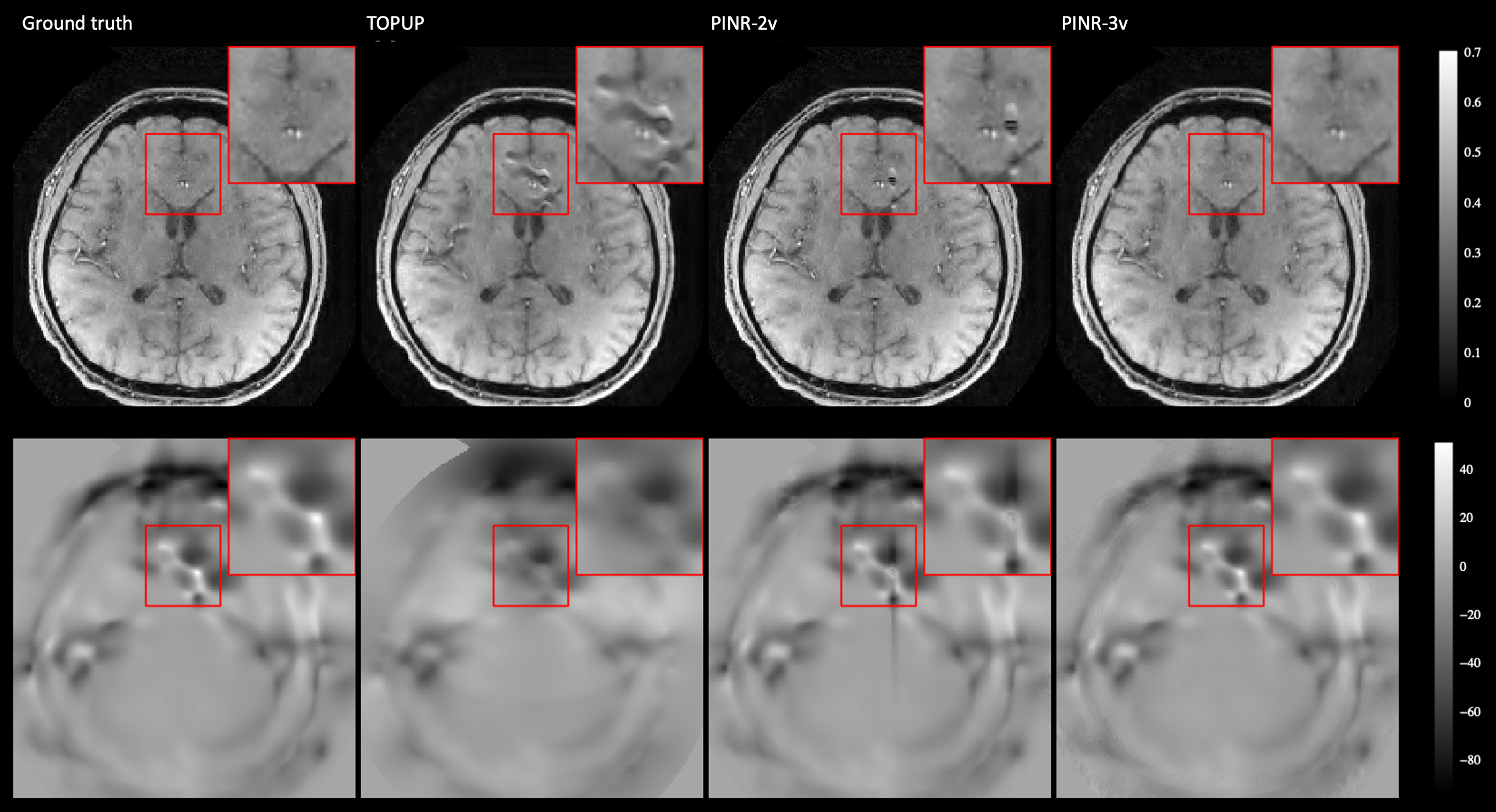}
    \caption{Comparison of two-view and three-view results. The first row presents the ground truth and corrected images, while the second row shows the corresponding B0 field estimates. The red box highlights the region of interest (ROI).}

    \label{fig:rot2vs3}
\end{figure}
% Figure~\ref{fig:rot2vs3} illustrates a more challenging case, where a high peak in the B0 map within the ROI severely distorts the image. With only two opposite views available, the correction is more difficult since distortions together with aliasing artifacts align along the same phase encoding direction. This limitation is evident in the TOPUP results, which fail to restore the region, leaving unresolved tissue compression in the ROI and producing an oversmoothed B0 estimation. PINR with two views captures more structural details in B0 than TOPUP but struggles at the high peak region, leading to local misestimation. In contrast, with three-view joint B0 estimation and image reconstruction, the additional rotated-view information helps disentangle distortions more effectively. The high peak in B0 is accurately estimated, resulting in improved image restoration and better-preserved anatomical structures. These results demonstrate the advantage of leveraging multiple views in PINR for handling severe distortions that are difficult to resolve using conventional blip-up/down methods.

Figure~\ref{fig:rot2vs3} shows a challenging case where a high B0 peak in the ROI severely distorts the image. With only two views, distortions and aliasing align along the same phase encoding direction, making correction difficult. TOPUP fails to restore the region, leaving unresolved tissue compression and oversmoothed B0 estimation. PINR-2v captures more structural details but struggles at the high peak, causing local misestimation. In contrast, PINR-3v leverages additional rotated views to better disentangle distortions, accurately estimating the B0 peak and preserving anatomical structures. These results highlight the benefits of multi-view PINR for handling severe distortions beyond conventional blip-up/down methods.

\section{Discussion and Conclusion}
We proposed PINR, a joint B0 estimation and image reconstruction framework using INRs for distortion correction in EPI images. Unlike conventional methods relying on explicit field maps or blip-up/down acquisitions, PINR optimized B0 and the distortion-free image within a physics-informed framework. Rotated-view EPI acquisitions helped distribute distortions across multiple orientations, improving B0 estimation and reconstruction, while INRs naturally handled off-grid multi-view data, making them well-suited for rotational sampling. Our experiments showed that PINR outperformed TOPUP across multiple metrics. Currently, the method operates slice-wise, limiting inter-slice consistency and efficiency. Additionally, validation was performed on simulated B0 data due to challenges in acquiring high-resolution field maps. More complex field imperfection such T2 blurring and eddy current are also not discussed. Future work will address these limitations with a 3D extension and in-vivo evaluation. Our results highlight the potential of INRs for MRI field imperfection correction, advancing physics-informed deep learning in medical imaging.

%% removed for anonymized MICCAI 2025 submission.
    
    % The following acknowledgement and disclaimer sections should be removed for the double-blind review process.  
    % If and when your paper is accepted, reinsert the acknowledgement and the disclaimer clause in your final camera-ready version.

\begin{credits}
\subsubsection{\ackname} This research was partially supported by European Research Council (Deep4MI project, Grant Agreement Nr.884622) and National Institutes of Health research grants (R01HD114719, K99EB035178, R01MH116173). Wenqi Huang’s research visit to Stanford University was also partly funded by TUM Graduate School Internationalization Support Grant.

\subsubsection{\discintname}
% It is now necessary to declare any competing interests or to specifically
% state that the authors have no competing interests. Please place the
% statement with a bold run-in heading in small font size beneath the
% (optional) acknowledgments\footnote{If EquinOCS, our proceedings submission
% system, is used, then the disclaimer can be provided directly in the system.},
% for example: 
The authors have no competing interests to declare that are
relevant to the content of this article. 
% Or: Author A has received research
% grants from Company W. Author B has received a speaker honorarium from
% Company X and owns stock in Company Y. Author C is a member of committee Z.
\end{credits}

\bibliographystyle{splncs04}
\bibliography{references}
\end{document}